\documentclass[twocolumn]{emulateapj}

\usepackage{latexsym,amssymb}
\usepackage{amsmath}
\usepackage{natbib}
\usepackage{graphicx,xcolor}
\usepackage[colorlinks=true,linkcolor=blue,citecolor=blue]{hyperref}

\newcommand{\Sim}{\mathord{\sim}}
\definecolor{darkred}{rgb}{0.7,0.1,0.1}

\slugcomment{Draft \today}

\shorttitle{Helium Detonations}

\shortauthors{Holcomb et al.}

\begin{document}
\author{Cole Holcomb\altaffilmark{1}, James Guillochon\altaffilmark{1}, Fabio De Colle\altaffilmark{1,2}, and Enrico Ramirez-Ruiz\altaffilmark{1}}
\altaffiltext{1}{TASC, Department of Astronomy and
  Astrophysics, University of California, Santa Cruz, CA
  95064}
   \altaffiltext{2}{Instituto de Ciencias Nucleares, Universidad Nacional Aut{\'o}noma de M{\'e}xico, A. P. 70-543 04510 D. F. Mexico}
   
\title{Conditions for Successful Helium Detonations in Astrophysical Environments}

\begin{abstract} 

Several models for type Ia-like supernovae events rely on the production of  a self-sustained detonation
powered by nuclear reactions. 
In the absence of hydrogen, the fuel that powers
these detonations typically consists of either pure helium (He) or a mixture of carbon and oxygen (C/O).
Studies that systematically determine the conditions required to initiate detonations in C/O material exist, but until now no analogous investigation of He matter has been conducted.
We perform one-dimensional reactive hydrodynamical simulations at a variety of initial density and temperature combinations and find critical length scales for the initiation of He detonations that range between 1 -- $10^{10}$ cm. 
A simple estimate of the length scales over which the total consumption of fuel will occur for steady-state detonations is provided by the Chapman-Jouguet (CJ) formalism. Our initiation lengths are consistently smaller than the corresponding CJ length scales by a factor of $\Sim 100$, providing opportunities for thermonuclear explosions in a wider range of low-mass white dwarfs (WDs) than previously thought possible.
We find that virialized WDs with as little mass as 0.24 $M_\odot$ can be detonated, and that even less massive WDs can be detonated if a sizable fraction of their mass is raised to a higher adiabat. 
That the initiation length is exceeded by the CJ length implies that certain systems may not reach nuclear statistical equilibrium within the time it takes a detonation to traverse the object. 
In support of this hypothesis, we demonstrate that incomplete burning will occur in the majority of He WD detonations and that $^{40}$Ca, $^{44}$Ti, or $^{48}$Cr, rather than $^{56}$Ni, is the predominant burning product for many of these events. 
We anticipate that a measure of the quantity of the intermediate mass elements  and  $^{56}$Ni produced in a helium-rich thermonuclear explosion can potentially be used to constrain the nature of the progenitor system. 

\end{abstract}


\section{Introduction}

The study of type Ia supernovae (SN Ia) occupies a critical position with regard to our understanding of the Universe. SN Ia have already delivered powerful insights into several areas of physics, especially cosmology and nucleosynthesis. Yet, despite their immense value to astrophysics, how SN Ia actually explode remains a vexing problem.  It has been generally accepted that these violent explosions are the result of the thermonuclear disruption of compact carbon and oxygen (C/O) white dwarf (WD) stars that have neared, or exceeded, the Chandrasekhar mass limit. Beyond that, however, little is certain. For instance, the mechanism(s) by which SN Ia progenitors reach their critical states is currently unknown, and more fundamentally, we remain ignorant as to what these critical conditions are.  Complicating the issue further, WDs only need to be close to the Chandrasekhar limit to be capable of auto-igniting at their centers, and external mechanisms can lead to the ignition of even less massive WDs. These mechanisms include collisions between WDs in dense stellar environments or in multiple stellar systems \citep{Rosswog:2009dx,Raskin:2009ie,Hawley:2012ea,Katz:2012tz}, tidal encounters with moderately massive black holes \citep{Luminet:1989wl,Rosswog:2008gc,RamirezRuiz:2009gw,Rosswog:2009gg,Haas:2012ci}, mergers of WDs with neutron stars (NS) or stellar mass black holes \citep{Fryer:1998kw, Fryer:1999gq, Lee:2007em,Metzger:2012ge,Fernandez:2013hb}, and the accretion of dense material from a WD companion \citep{Nomoto:1982hy,Woosley:1986hh,Bildsten:2007eh,Fink:2007hi,Fink:2010jt,Guillochon:2010bp,Shen:2010gb,Woosley:2011fu,Dan:2011ff,Dan:2012bc,Sim:2012kn}. SN Ia initiation and the diversity of possible outcomes once ignition occurs are therefore pressing issues to the astrophysics community that call for further study.

\begin{figure}[]
\centering\includegraphics[width=\linewidth,clip=true]{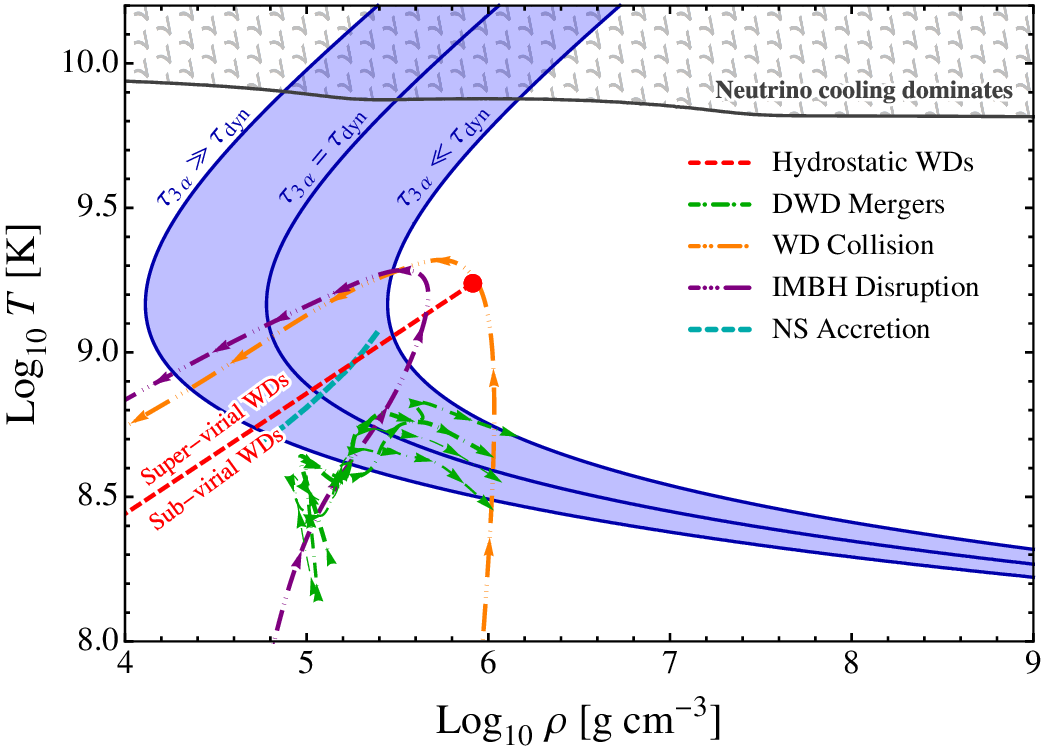}
\caption{The $\rho-T$ parameter space relevant to He ignition. The blue-shaded region encompasses the $\rho-T$ values for which a He detonation might occur, and is defined by the constraint $0.1< \tau_{\rm dyn}/\tau_{3\alpha}<10$. The $\nu$-filled region above approximately eight billion Kelvin shows where the energy losses due to neutrino emission \citep[pair annihilation, plasma, photoneutrino, recombination, and bremsstrahlung processes,][]{Itoh:1996bo} dominate over the energy generation due to triple-$\alpha$ reactions \citep{Khokhlov:1986ve,Rosswog:2008gc}. The red-dashed line represents the $\rho-T$ trajectory of virialized $n=3/2$ polytropes (Hydrostatic He WDs) up to $0.45 M_{\odot}$ in total mass; the core density and temperature for a $0.45 M_{\odot}$ WD are denoted by the solid red dot. The green-dashed lines show the time evolution of double white dwarf (DWD) mergers with a $0.4 M_{\odot}$ donor and accretors of $0.8 M_{\odot}$, $0.9 M_{\odot}$, $1.0 M_{\odot}$, and $1.1 M_{\odot}$, where the plotted line thickness increases with donor mass, which were produced using the simulations presented in \citet{Dan:2012bc}. The orange-dashed line corresponds to the head-on collision of two $0.4M_{\odot}$ He WDs, and were obtained from the simulations of \citet{Rosswog:2009dx}. The purple-dashed line exhibits the parameters of a $0.2 M_{\odot}$ WD as it is tidally disrupted by a $5000M_{\odot}$ BH at $\beta=5$, and is taken from \citet{Rosswog:2009gg}. The dashed cyan line shows the trajectory of the accretion disc formed by the merger of a $0.3 M_\odot$ WD and a $1.4 M_\odot$ NS calculated in \citet{Metzger:2012ge}. Arrows are used throughout to indicate time evolution.}
\label{fig:crescent}
\end{figure}

Several models have been proposed for the sequence of occurrences that lead to SN Ia-like events, but all share the common feature that a self-sustained detonation powered by nuclear reactions propagates through a large fraction of the interior of the host object (or objects). In the absence of hydrogen, the fuel that powers this detonation consists of either pure helium (He) or the products of He burning, typically a mixture of carbon and oxygen. This detonation must be triggered by an increase in density, temperature, or both. The conditions for the initiation of detonations is a topic that has been discussed at length in the context of C/O burning \citep{Arnett:1994ev,Niemeyer:1997dg,Ropke:2007dd,Seitenzahl:2009jf}, but only to a limited extent for the ignition of pure He \citep{Khokhlov:1986ve,Khokhlov:1989vg}. In this paper, we present a parameter space study to find the critical temperature gradient length scales necessary to successfully initiate a detonation in He media for a given combination of density and temperature.

A detonation will occur when the nuclear timescale is on the order of the local dynamical timescale. However, the definition of both of these timescales is ambiguous; the effective nuclear timescale depends on a number of reactions with different  rates, and the effective dynamical timescale depends on the size of the region being heated. Furthermore, the heated region is most likely not uniform in either composition or temperature at any particular time, and even less so over the timescale in which the heating occurs. We can analytically estimate the length scale $\xi$ that must be heated to produce a detonation by calculating the speed at which the detonation will travel $u$ and then multiplying by the timescale of the corresponding nuclear reactions $\tau_{\rm nuc}$, $\xi = u \tau_{\rm nuc}$. Both $u$ and $\tau_{\rm nuc}$ can be calculated by using simple one-dimensional models of the detonation in combination with a nuclear network \citep{Khokhlov:1989vg}. This estimate is an upper limit on the size required, as heating a region larger than this scale would result in an overdriven detonation that would eventually slow to the steadily propagating solution. However, while this method can establish the length scale of the resulting detonation structure, it is unclear if a region smaller than $\xi$ may be capable of yielding a detonation, especially given that the initial conditions are unlikely to resemble the assumed steady-state detonation structure. For instance, the hydrodynamical simulations of He detonations in WD atmospheres of \citet{Townsley:2012gz} show that steadily propagating weak detonations can often strengthen towards a steady-state solution, though only after the detonation has traveled some distance.

To determine the critical temperature gradient length scale for a given density and floor temperature, while simultaneously resolving the initiation region and subsequent shock wave front, we perform one-dimensional reactive hydrodynamical calculations that begin with a hot spot in a He medium and check if they lead to detonations. Our measurement of these critical length scales allow further constraints to be placed on SN Ia models as well as other more exotic thermonuclear explosions involving the disruption of He WDs. For many systems, we find that the spatial scales relevant for the initiation of these He detonations are unresolved in even the most computationally expensive multidimensional simulations, with a continuum of critical lengths $\xi_{\rm crit}$ that ranges from 1 cm to $10^{10}$ cm.

This paper is structured as follows. In Section \ref{sec:con}, we present the SN Ia initiation models in question and the density and temperature conditions in which they subsist. In Section \ref{sec:met} we describe the numerical methods and initial conditions used for the study and present the results. In Section \ref{sec:dis} we discuss the results and their application to thermonuclear  explosion models.

\section{Relevant Conditions}\label{sec:con}

White dwarfs, the end point in the evolution of the vast majority of stars in the universe, are extremely common, accounting for about 65\% of the total heavy element abundance \citep[i.e. heavier than He,][]{Fukugita:2004bf}. 
Because WDs experience no mass loss through single stellar evolution, their material can only be ejected through dynamical interactions with other objects and/or by the rapid release of nuclear binding energy.
Although at present the reservoir is dominated by C/O WDs, which possess a thin layer of He \citep{Lawlor:2006fp}, a non-negligible fraction of the mass lies within pure He WDs \citep[with masses in the range of 0.17--0.45$M_{\odot}$, $\Sim 6$\% of WDs,][]{Brown:2010fd,Kleinman:2013hn} and hybrid WDs consisting of small C/O cores and thick He envelopes with total masses of $\Sim 0.5M_{\odot}$ \citep{Iben:1985bn,Nelemans:2001ia}.
Pure He WDs cannot be the result of the evolution of single stars within the lifetime of our Galaxy; these WDs can only be produced via interacting binary systems \citep{Nelemans:2001ia,Nelemans:2001gy,Rappaport:2009kg} and approximately half of all WDs in close binaries are expected to be either He-core or hybrid WDs \citep{Ruiter:2011ez}.
In addition to systems in which both stars are WDs, binary systems in which the accretor is either a neutron star or a black hole are also possible. As a result, the conditions under which He may be ignited are diverse, with a variety of accretion rates, initial entropies, and surface gravities.

In Figure~\ref{fig:crescent} we present the density-temperature parameter space relevant to He ignition. If the timescale on which the He material can react, its dynamical timescale $\tau_{\rm dyn}= (G{\rho})^{-1/2}$ s, is much shorter than the burning timescale $\tau_{3\alpha}=9.0\times 10^{-4} T_9^3 \exp{(4.4/T_9)}\rho_6^{-2}$ s \citep{Khokhlov:1986ve}, the material can expand rapidly enough to quench burning ($T=10^9T_9$\;K and $\rho=10^6\rho_6$ g\;cm$^{-3}$). This is because in a pure He environment the combustion rate is limited by the triple-$\alpha$ reaction. Appreciable burning will therefore only take place if $\tau_{3\alpha}\lesssim \tau_{\rm dyn}$. This comparison of timescales can be used as a simple estimate for whether or not substantial He burning can occur in particular physical systems, but the onset of dynamical burning does not guarantee that a self-sustained detonation will form, as the conditions for detonation are non-trivial \citep{Khokhlov:1989vg}. However, we do not expect that detonations will occur in systems where $\tau_{3\alpha} \gg \tau_{\rm dyn}$, because burning would likely  be quenched by expansion. Above temperatures of approximately $8 \times 10^{9}$ K, energy loss by neutrino emission begins to overtake the energy production due to triple-$\alpha$ reactions (the gray-shaded region in Figure~\ref{fig:crescent}). We have therefore restricted our simulations to the region in which $\tau_{3\alpha}\le 10\tau_{\rm dyn}$ and in which the relation $|\dot{E}_{\nu}|<\dot{E}_{3\alpha}$ holds, where $\dot{E}_{\nu}$ includes contributions from pair annihilation, plasma, photoneutrino, recombination, and bremsstrahlung processes \citep{Itoh:1996bo}. For comparison, we have included in Figure~\ref{fig:crescent} the trajectories for virial hydrostatic He WDs; that is, we have calculated the parametric density-temperature relation as a function of radius for WDs that are raised to temperatures such that their degeneracy and radiation pressures are roughly equal. To model these WDs, we solve a polytropic equation of state with $n=3/2$ and with masses up to $0.45M_{\odot}$.
In addition, we also display in Figure~\ref{fig:crescent} the trajectories for double WD mergers, collisions of WDs,  tidal compression events of WDs orbiting black holes, and the disruption of He WDs orbiting  neutron stars. These trajectories all fall within the region of interest, and thus are suitable candidates for systems in which successful He detonations can occur.

\section{Conditions for Successful Helium Detonations}\label{sec:met}

\subsection{Numerical Methods and Initial Conditions}

To study the detonation of He, we have carried
out a set of 1D cartesian simulations using the adaptive mesh refinement 
code \emph{Mezcal}.
The code has been tested extensively against standard test problems, 
and has been used to simulate magnetohydrodynamic
jets \citep[e.g.,][]{deColle:2006bf,DeColle:2008gd} and 
gamma-ray burst afterglows \citep{DeColle:2012gr,DeColle:2012cy}. In \emph{Mezcal}, the equations of hydrodynamics are integrated in time using an explicit
second-order Runge-Kutta method, while the fluxes are calculated 
using the HLL method, with a second-order spatial 
reconstruction of the primitive variables at the cell interfaces.
A thirteen element ``$\alpha$-chain'' nuclear network, which includes all alpha process elements from $^{4}$He to $^{56}$Ni \citep{Timmes:2000ft} and energy losses due to neutrino cooling, is integrated in time by solving the system of equations
\begin{equation}
  {\partial \rho_i\over\partial t}+\nabla \cdot (\rho_i \vec v)= \Gamma_i,
  \label{eq:hd}
\end{equation}
where $\rho_i$ is the mass density of the element $i$ (with $i = 1, \dots, 13$), $\vec v$ is the velocity vector and $\Gamma_i$ is the total reaction rate for the element $i$. The system of equations \ref{eq:hd} are coupled to the hydrodynamics by an operator splitting method \citep{LeVeque:2002uu}. We first integrate equations \ref{eq:hd} by setting $\Gamma_i=0$, and then we use the new $\rho_i$ values to integrate the set of ordinary differential equations \citep{Timmes:2000wm}
\begin{equation}
  {d\rho_i\over dt} = \Gamma_i.
  \label{eq:ti}
\end{equation}
Finally, the original internal energy within a cell $e_0$ is updated based on the energy loss and gain due to the nuclear reactions and the neutrino cooling to obtain the new internal energy $e_1$.

To properly couple the hydrodynamics with energy injected by burning, and removed by neutrinos,
the timestep is limited such that the energy change is less than 0.5\% of the internal energy for all cells. This is achieved by setting the time step $\Delta t_{\rm nuc}^n = 5 \times 10^{-3}  \Delta t^{n-1}  e_0/ \lvert e_1-e_0\rvert$, where $e_0$ and $e_1$ are the thermal energies before and after the update from the nuclear routine.
Otherwise, the timestep is set by the Courant condition, defined as the minimum among all the cells of $\Delta x/(v+c_{\rm s})$, 
where $\Delta x$, $v$ and $c_{\rm s}$ are the size of the cell, the velocity, and
the sound speed respectively. As in other hydrodynamical studies, both the energy release calculated by a reduced network and the fraction of $^{56}$Ni produced can vary by $\Sim 30$\% when compared to a 489 isotope network \citep{Timmes:2000ft}, or when using a coupling constant that is larger than what is used in this study \citep{Hawley:2012ea}. Therefore, while we expect that this reduced network should capture the general nucleosynthetic trends, the normalization of these trends may differ from what is presented here.

In the \emph{Mezcal} code, a basic Cartesian grid is
built at the start of the simulation, and it is refined based
on the initial conditions and the subsequent evolution of the
flow \citep[see][]{DeColle:2012gr}. In the simulations presented
here, we use a refinement (and derefinement) criterion based on the ensuing density
gradients. In particular, when the fractional increase in density, computed by comparing a cell
with its two neighbor cells, is larger (smaller) than 0.5 (0.05), the cell is refined (derefined) 
and two new cells are created (or the old cells are eliminated).

For our study we initialized linear temperature gradients (``hot spots'') of a particular size $\xi$ in a pure He environment of density $\rho$, where the maximum temperature $T_{\max}$ occurs at $x = 0$. We assume one-dimensional cartesian geometry with reflecting boundary conditions at $x = 0$ and outflow boundary conditions at $x = x_{\max}$. For consistency, we always chose the temperature floor $T_{\rm 0} = 10^{-2} T_{\max}$. We employed the method of bisection to determine the minimum size $\xi_{\rm crit}$ of the hot spot that resulted in a self-sustained detonation for a given $T_{\max}$ and $\rho$.

As a test of the implementation of the nuclear network module in the \emph{Mezcal} code, we first reproduced the resolution study for C/O detonations as presented in section 3.3 of \citet{Seitenzahl:2009jf}, and also did not find convergence until the minimum grid cell size $\Delta x_{\min} \lesssim 2$ cm. We then performed a resolution study for a pure He composition, with $T_{\max} = 10^{9}$ K, $T_{0} = 10^{7}$ K, and $\rho = 10^{6}$ g cm$^{-3}$, and found convergence for $\Delta x_{\min} \lesssim 10^{4}$ cm, which is approximately equal to the estimate of the width of the burning region presented in Table I of \citet{Khokhlov:1986ve}.

\begin{figure*}[htb]
\centering\includegraphics[width=0.9\linewidth,clip=true]{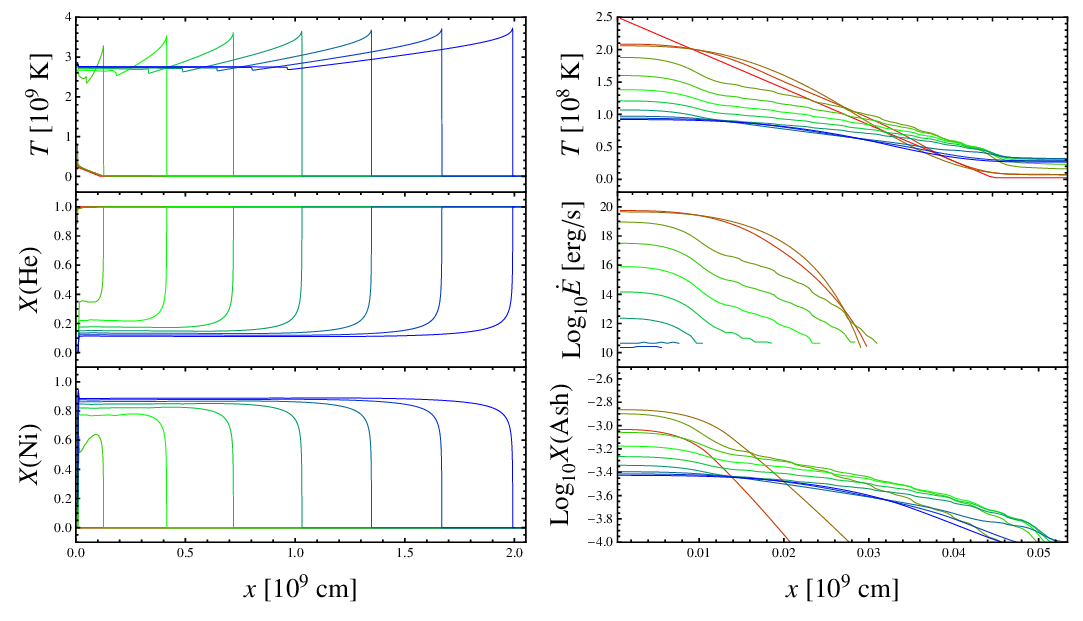}
\caption{A Successful Detonation Versus A Failed Detonation. Depicted on the left is the time evolution of a successful (critical size) He detonation, initialized with $T_{\rm max} = 0.25\times 10^9$ K, $\rho = 10^6$ g cm$^{-3}$, and $\xi_{\rm crit} = 0.11\times 10^9$ cm. The hot-spot quickly develops into a full blown shock-driven detonation. The shock couples with the rapid fusion of approximately 90\% of the initial He fuel into nickel, producing peak temperatures of $\Sim 4 \times 10^9$ K.  On the right is displayed a failed detonation in which the size has been reduced to $\xi=0.48\times 10^8$ cm, a reduction of roughly a factor of two from the critical size; the temperature and density are unchanged from the left panel. This initial temperature gradient is unable to form a shock. Despite the initially positive energy generation rate, the products of the triple-$\alpha$ reactions (collectively defined as {\it Ash}) are nearly negligible. In both panels the initial state is indicated in red, intermediate states in green, and latest states in blue.}
\label{fig:profiles}
\end{figure*}

We determined which simulations successfully produced detonations by identifying shocks that propagated at a near-constant velocity, indicating that the were powered by the continual release of energy through sustained nuclear reactions. In Figure~\ref{fig:profiles} we show the results of two different simulations, where the initial conditions differ only in the extent $\xi$ of their hot spots. In the left column of Figure~\ref{fig:profiles}, the temperature gradient quickly evolves into a fully-formed shock wave, rapidly converting the majority of the He fuel into Ni (bottom two panels). The right column shows a case where $\xi$ has been reduced by a factor of $\Sim 2$. In this case, the burning that is initiated is unable to outpace the hydrodynamic expansion of the fluid, even though triple-$\alpha$ reactions initially inject a significant amount of energy into the system. The boundary between successful and unsuccessful detonation is sometimes signified by the inability to produce a shock. This is in contrast to what \citet{Seitenzahl:2009jf} found for the initiation of C/O detonations; they found a shock is produced even in sub-critical conditions.

\begin{figure}[htb]
\centering\includegraphics[width=\linewidth,clip=true]{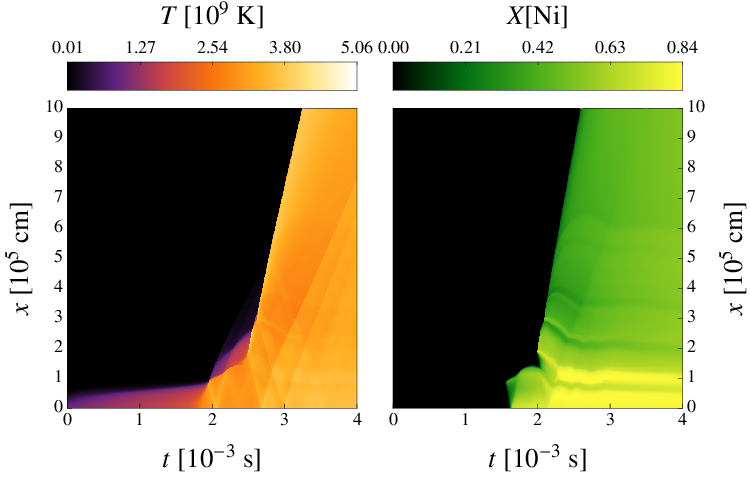}
\caption{Time evolution of temperature and nickel fraction for a successful detonation. The above panels display a successful detonation initialized with $T_{\rm max}=10^9$ K, $\rho=5\times 10^6$ g cm$^{-3}$, and $\xi_{\rm crit}=6.1\times 10^4$ cm. On the left it is shown that the temperature reaches a peak of $T=5.06\times 10^9$ K for a brief moment during the ignition of the shock wave. The detonation-front subsequently relaxes to a temperature of approximately $T\approx 4\times 10^9$ K. Three ignition phases occur before the shock obtains stability; multiple ignitions are characteristic of high density ($\rho_6 \gg 1$) detonations. The ignitions produce shocks in both directions; reflections of the backward traveling shocks due to the left boundary are clearly visible. In the right panel the nickel production is shown. The ignition phases of the detonation convert up to nearly $85\%$ of the He fuel into nickel, while the stable shock settles to a conversion of about $50\%$. The large abundance of nickel in the hot-spot region is not a general result within the parameter space (c.f. Figure~\ref{fig:ti}).}
\label{fig:twod}
\end{figure}

In Figure~\ref{fig:twod}, we show the time evolution of $T$ and $X_{\rm Ni}$ for a successful detonation. For $\rho > 5 \times 10^{6}$ g cm$^{-3}$ it is common for the initial conditions to lead to multiple runaways that only produce a stably-propagating detonation through their mutual interactions. These multiple runaways come as the result of the strong density dependence of the triple-$\alpha$ reaction. The expansion of the hot spot introduces upstream (pre-shock) perturbations to the density, which can potentially experience dynamical burning if the temperature in these regions is $\Sim 10^{9}$ K. Each runaway produces a pair of left- and right-traveling shock waves that interact with the hot spot as it expands, and collide both with shocks produced from other runaway regions, and with the reflecting boundary condition at $x = 0$. While this behavior is somewhat pathological, our criteria for the boundary between success and failure is still applicable, as we ensure that these multiple interactions eventually produce a single shock that propagates at a constant speed.

\begin{figure}[t]
\centering\includegraphics[width=\linewidth,clip=true]{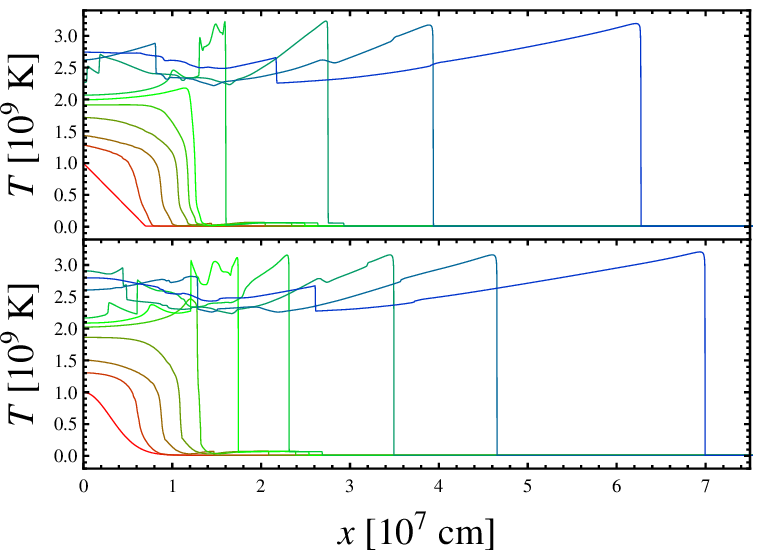}
\caption{Linear versus Gaussian initial temperature gradients. The above detonations are initialized with $T_{\rm max}=10^9$ K and $\rho =10^6$ g cm$^{-3}$. The top panel displays a linear hot spot with critical radius $\xi_{\rm crit} = 7.0\times 10^6$ cm, and for comparison, the bottom panel displays a Gaussian hot spot with critical radius $R_{\rm crit}=4.3\times 10^6$ cm. It is evident that the only appreciable difference between the two is that the detonation with the Gaussian profile reaches the ignition point earlier than the linear profile. The times displayed are not uniformly distributed. }
\label{fig:gaussian}
\end{figure}

The choice to use a linear temperature profile is admittedly arbitrary, so we determined the critical size for a Gaussian temperature distribution and compared the results to a linear temperature profile with the same $T_{\max}$, $T_{0}$ and $\rho$. Figure~\ref{fig:gaussian} shows the time-evolution of a critical detonation for these two cases where $\xi = \xi_{\rm crit}$ (for the linear case) or $R = R_{\rm crit}$ (for the Gaussian case, where $R$ is equal to the standard deviation). While the delay between the start of the simulation and the time at which a self-propagating detonation forms is different between the two simulations, the resulting detonation structure is almost identical.

\begin{figure*}[t!]
\centering\includegraphics[height=5.75cm,clip=true]{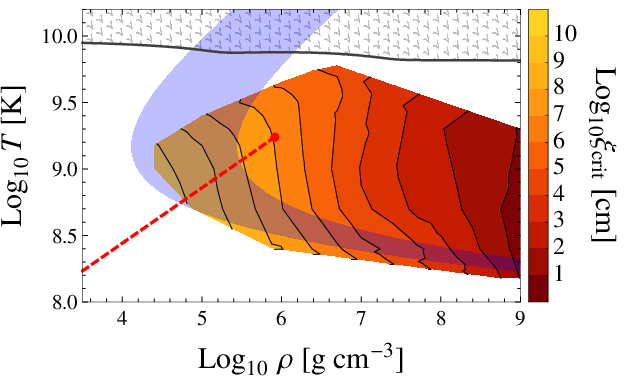}\includegraphics[height=5.75cm,clip=true]{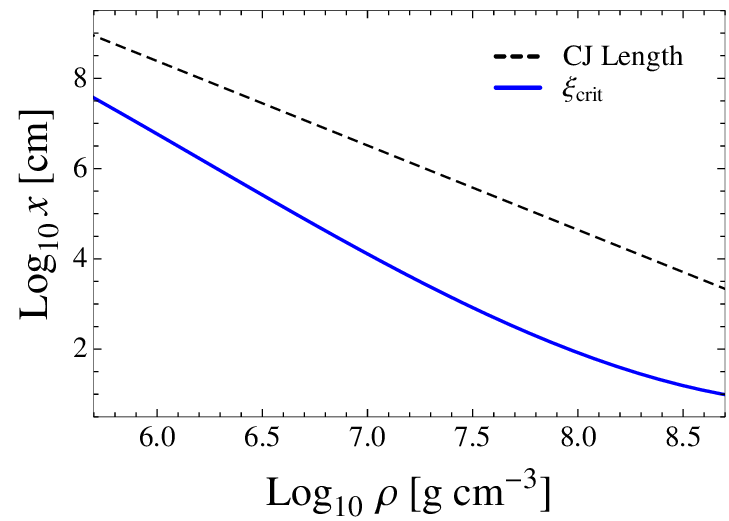}
\caption{Critical sizes for He detonation. The left-hand panel shows the minimum size scale $\xi_{\rm crit}$ for a linear temperature gradient as a function of density $\rho$ and peak temperature $T_{\rm max}$. The $\tau_{3\alpha}\sim \tau_{\rm dyn}$ and $|\dot{E}_{\nu}|>\dot{E}_{3\alpha}$ shaded zones, along with the virial WD line, have been carried over from Figure~\ref{fig:crescent}. $\xi_{\rm crit}$ ranges from approximately $1$ cm to $10^{10}$ cm, and is weakly dependent on temperature. In the right-hand panel, we present a comparison of the critical size $\xi_{\rm crit}$ against the width of the CJ burning zone as a function of density. The dashed black line represents the $l_{\rm CJ}$, which is equal to the product of the burning timescale and the detonation velocity \citep{Khokhlov:1989vg}, and the solid blue line represents $\xi_{\rm crit}$ assuming an initial temperature $T_{\rm max}=10^9$ K (corresponding to an upstream temperature of $T_{0} = 10^7$ K).}
\label{fig:sizes}
\end{figure*}

\begin{figure*}[t]
\centering\includegraphics[width=\linewidth,clip=true]{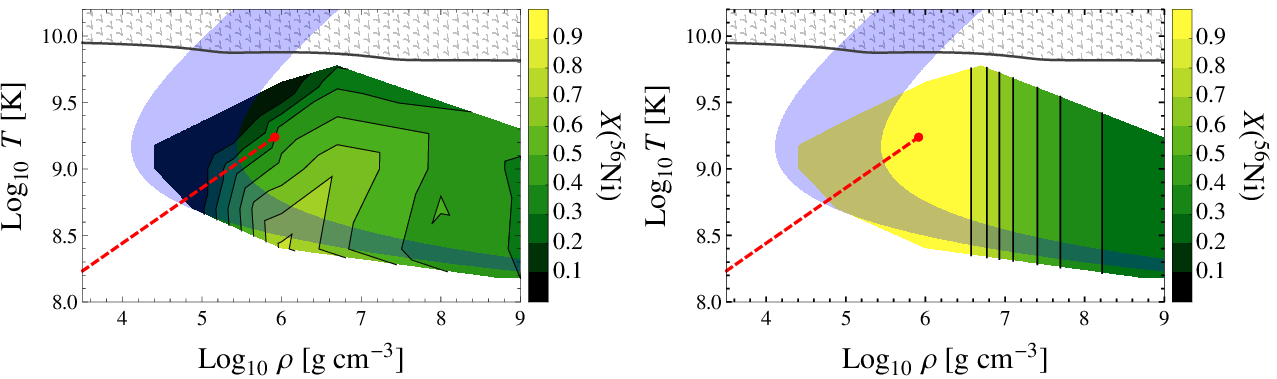}
\caption{Simulated versus NSE nickel abundances. The left panel displays the average fraction of nickel produced by a critical detonation within the zone delimited by $3\xi_{\rm crit} \le x \le 4\xi_{\rm crit}$ when the shock front reaches $x=4\xi_{\rm crit}$. These measured values range from zero to $\Sim 0.7$. In the right panel we have calculated the NSE values of nickel abundance under the relevant conditions that correspond to the $\rho$ and $T_{\rm max}$ points of the left panel (i.e. $\rho$ and $T_0=10^{-2}T_{\rm max}$).}
\label{fig:ni}
\end{figure*}

\begin{figure*}[t!]
\centering\includegraphics[width=\linewidth]{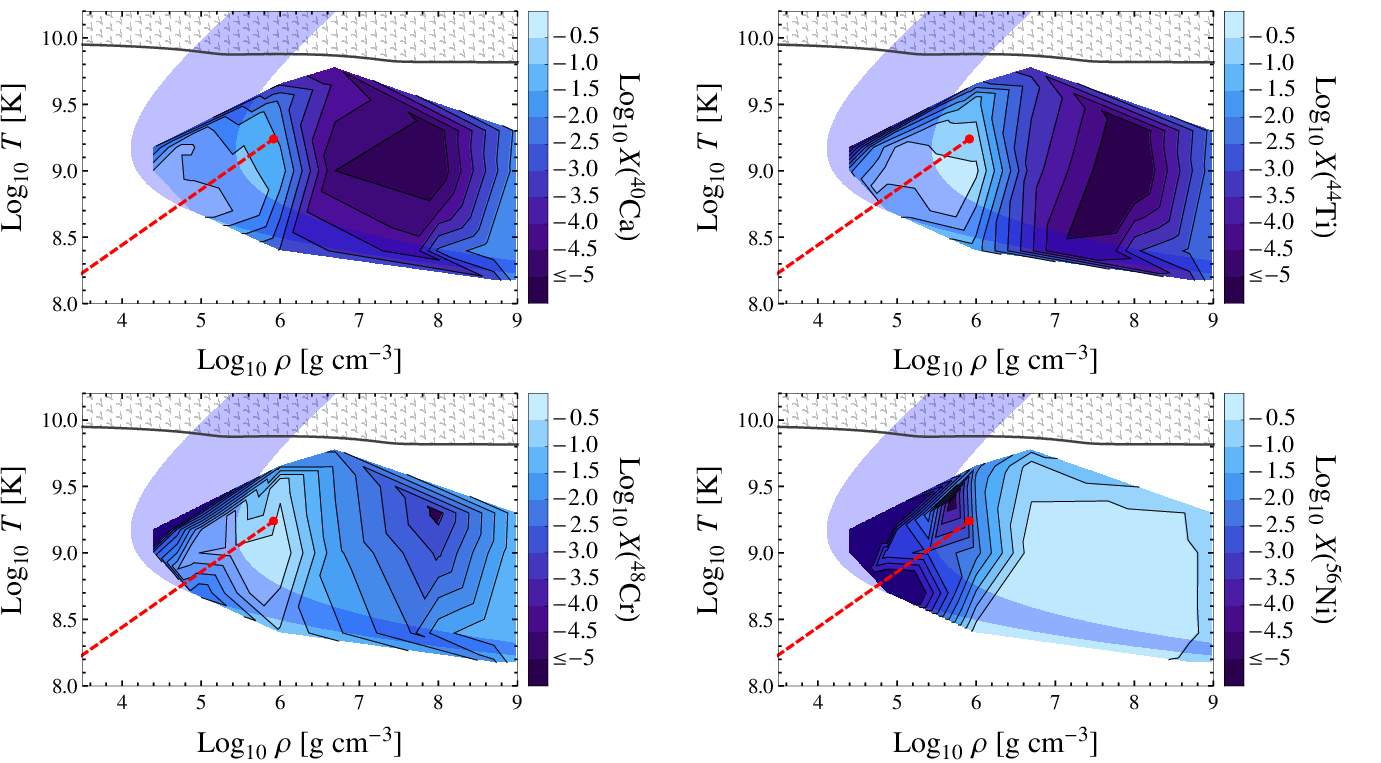}
\caption{Titanium, nickel, calcium, and chromium yields in the hot spot zone. The above panels display the average abundances of calcium (top left), titanium (top right), chromium (bottom left), and nickel (bottom right) within the hot spot zone ($0\le x \le \xi_{\rm crit}$) as a function of $\rho$ and $T_{\rm max}$. These measurements were taken when the shock front crossed $x=4\xi_{\rm crit}$.}
\label{fig:ti}
\end{figure*}

\subsection{Critical Size Estimates}\label{subsec:critsize}

Past efforts in determining minimal length scales
for the initiation of detonations have been concerned with C/O WDs  
\citep{Arnett:1994ev,Niemeyer:1997dg,Ropke:2007dd,Seitenzahl:2009jf}. In all of the aforementioned works, these critical sizes were determined through the use of one-dimensional hydrodynamical simulations, but while these studies have been performed by numerous groups for C/O mixtures, they have never been performed for pure He. In \citet{Moll:2013vq} the critical size required for the successful detonation of a helium envelope with mass $0.045 M_{\odot}$ on the surface of a $1 M_{\odot}$ C/O WD was recently determined through the use of multidimensional simulations. However, a systematic determination of the critical length scales for all possible combinations of core and envelope mass would be computationally prohibitive in multiple dimensions.

In the left panel of Figure~\ref{fig:sizes}, we show the minimum size scale $\xi_{\rm crit}$ for combinations of $\rho$ and $T_{\rm max}$ that successfully resulted in a detonation. These scales range from $\Sim 1$ cm at $\Sim 10^{9}$ g cm$^{-3}$ to $10^{10}$ cm at $10^{5}$ g cm$^{-3}$, with a relatively weak dependence on $T$ at each $\rho$. Overlayed on this panel are the same regions highlighted in Figure~\ref{fig:crescent}, where the red line again shows the virial relation for WDs. Within the blue crescent (where dynamical burning occurs) and along the WD relation, the minimum scale ranges from $10^{8}$ -- $10^{10}$ cm, similar to the range of WD sizes. As shown in Figure~\ref{fig:sizes}, the critical $T$ necessary to produce a detonation at a given size $\xi$ is comparable to the virial temperature of the WDs themselves, and thus these conditions can potentially be achieved through dynamical processes that are driven by the WD's own gravity (e.g. collisions or the acquisition of material from a companion). The He shell detonations of \citet{Moll:2013vq} are consistent, within a factor of 2-3, with our results; for a spherical cap initiation zone of $\Sim 7\times 10^5$ g cm$^{-3}$ in density and a peak temperature of $2\times 10^9$ K, the necessary size for successful detonations was found to be on the order of $10^7$ cm \citep[see ``model D" of][]{Moll:2013vq}.

It is apparent from the left panel of Figure~\ref{fig:sizes} that $\xi_{\rm crit}$ is primarily a function of density, with only a slight dependence on temperature as $T$ approaches the $\nu$ cooling limit and where the dynamical and burning timescales are comparable. In the right panel of Figure~\ref{fig:sizes} we show how the initiation scale varies as a function of density, where the peak temperature of the linear gradient $T_{\rm max}$ is assumed to be $10^{9}$ K (corresponding to an upstream temperature $T_0 =10^7$ K), and compare it to the Chapman-Jouguet (CJ) length $l_{\rm CJ} \equiv D\tau_{3\alpha}$, where $D$ is the CJ velocity for that particular density and $\tau_{3\alpha}$ is calculated using the corresponding downstream (post-shock) temperature \citep{Khokhlov:1989vg}. We find that the initiation scales at these densities are such that $l_{\rm CJ} \gtrsim 100 \xi_{\rm crit}$, implying that burning may not proceed to completion within the time it takes the detonation to traverse the system.

\subsection{Nucleosynthesis}\label{subsec:nucsynth}

In the simple CJ model of a detonation \citep{Chapman:1899ko,Jouguet:1905to}, the burning is assumed to be instantaneous across an infinitely-thin zone behind the shock, and the detonation propagates at a single speed, the CJ velocity $D$. In reality, the reactions that take place occur in a finite timespan, and thus the region over which the reactions complete has a finite size. Zel'dovich, von Neumann, and D\"{o}ring (ZND) \citep{Zeldovich:1940wi,vonNeumann:1942wx,Doring:1943tn} improved this by modeling a detonation as a one-dimensional leading shock wave moving at the detonation speed trailed by a reaction zone of finite width \citep{Fickett:1979vz}. This enables one to estimate a variety of length scales, including the region over which the bulk of the energy is injected, the distance over which the fuel is exhausted, and the distance over which the composition approaches nuclear statistical equilibrium (NSE) \citep{Timmes:2000jg}.

For systems in which the CJ and ZND scales are much smaller than the scale of the He fuel region, the burning is expected to yield a composition of material composed primarily of He and nickel, because the timescale for $\alpha$-capture is short compared to the triple-$\alpha$ timescale \citep{Khokhlov:1989vg}. If the burning is quenched before the detonation has traveled a few CJ lengths, the burning cannot proceed to completion, as is the case for systems in which the initiation scale is comparable to the local fuel scale. Alternatively, the material can expand, thereby quenching the burning before alpha particles have accumulated on each seed nucleon to produce $^{56}$Ni \citep{Townsley:2012gz}.

Our results are consistent with this picture. In the left panel of Figure~\ref{fig:ni} we calculate the fraction of $^{56}$Ni in the region immediately downstream to the detonation within our simulations shortly after the detonation reaches a steadily-propagating state. To do this, we average the nickel abundance within $3\xi_{\rm crit} \le x \le 4\xi_{\rm crit}$ at the moment the shock crosses $x=4\xi_{\rm crit}$. These measurements are compared to the expected fraction of $^{56}$Ni if the detonation is allowed to propagate a distance long enough for the downstream material to reach NSE (Figure~\ref{fig:ni}, right panel). We find that the two models produce radically different yields of $^{56}$Ni, indicating that the post-detonation state that is produced shortly after a detonation first forms is not representative of the state it would have reached if it was allowed to expand until NSE was achieved.

As described in Section \ref{subsec:critsize}, the initiation size $\xi_{\rm crit}$ is comparable to the size of the WD itself for $\rho \lesssim 10^{6}$ g cm$^{-3}$. Incidentally, this is the same region in which the fraction of $^{56}$Ni declines to the point that it is no longer the predominant species within the ash (Figure~\ref{fig:ti}, right panel. Here the titanium abundance is averaged over the hot spot zone, which we define to be $0\le x \le \xi_{\rm crit}$, at the time the shock reaches $x=4\xi_{\rm crit}$). For densities $10^{5} \lesssim \rho \lesssim 10^{6}$ g cm$^{-3}$, and for temperatures typical of adiabatically compressed WD material (red dashed line), the predominant species that is produced is $^{44}$Ti (Figure~\ref{fig:ti}, left panel.). This progression to heavier species in the ash as the density increases is similar to the result seen when He is burned at constant density and temperature \citep{Hansen:1971el}.

The stable decay product of $^{44}$Ti is $^{44}$Ca, and the production of this particular isotope has been associated with type II SNe. However, the amount of $^{44}$Ti produced per type II SN may not be sufficient to explain the amount required to produce the ratio of $^{44}$Ca/$^{40}$Ca observed in the Sun \citep{Timmes:1996gh}. Given the fact that greater than half of the initial He can be burned to $^{44}$Ti, He burning can produce vastly larger quantities of $^{44}$Ti than type II SNe, which on their own may not be able to explain observed abundance \citep{Woosley:1994ik,The:2006bu}. If the rate of WD collisions, disruptions, and SN Ia events involving a He WD are comparable to the type II rate, He burning on low-mass WDs may be the primary source of $^{44}$Ti in the universe.

\section{Discussion}\label{sec:dis}

We have presented one-dimensional results of the determination of critical (smallest) spatial scales for the gradient initiation of detonations in He matter using the reactive hydrodynamics code \emph{Mezcal}.  In particular, we have shown that the spatial scale $\xi_{\rm crit}$ required for the initiation of a detonation differs significantly from the CJ size, specifically, it is approximately two orders of magnitude smaller for a wide range of densities. That the initiation length $\xi_{\rm crit}$ is greatly exceeded by the CJ length implies that certain systems may not reach nuclear statistical equilibrium within the time it takes a detonation to traverse the He material. The existence of a small region in which a detonation can be initiated does not guarantee that a sufficient length of fuel is available to ensure NSE is eventually achieved. Therefore incomplete burning that terminates at an element lighter than $^{56}$Ni will occur in many physical systems in which the size of the fuel reservoir is comparable to $\xi_{\rm crit}$.
 
In astrophysical systems for which the state of the fluid is initially determined by a balance of pressure and gravitational forces, a compressive event will typically occur on the local dynamical timescale, which naturally results in the material being heated to the local virial temperature. Whether this action leads to a detonation depends on whether or not the region that is being compressed has a scale that is comparable to $\xi_{\rm crit}$. In systems where there is a prolonged period in which He experiences repeated compressions, e.g. mass transferring or merging binary systems, the criteria for detonation will be satisfied as soon as the length scale of the region being compressed is comparable to the initiation scale. These repeated compressions are seen in stably-accreting, unstably-accreting, and merging systems, and are driven by convective motions and/or fluid instabilities \citep{Fryer:1999gq,Shen:2009bx,Guillochon:2010bp,Fernandez:2013hb}.

Within an atmosphere in which the mass of the atmosphere is small compared to the mass of the host object, the density falls off exponentially, so material will only have approximately constant density over the scale height $H$. As $\xi_{\rm crit}$ is a strong function of density (Figure~\ref{fig:sizes}), the detonation is not expected to be sustained as it propagates upwards through such a steep density gradient, resulting in a ``blowout'' \citep{Townsley:2012gz} that quenches burning after propagating in the vertical direction a distance $\Sim H$. However, the detonation can continue to propagate in the horizontal direction, as the bulk conditions in the base of the atmosphere of a nearly-spherical body do not vary much with the position, resulting in a continual blowout of partially-burnt material as the detonation propagates around the host object.

We find that for densities typical of WDs, the predominant element that is produced from burning is not necessarily $^{56}$Ni, and is more commonly an intermediate $\alpha$-chain element such as $^{44}$Ti. Similar to the blowout case presented in \citeauthor{Townsley:2012gz}, the detonation is quenched in such systems after propagating a distance comparable to $\xi_{\rm crit}$. Therefore, the results of our simulations may be used to approximate the nucleosynthetic yields resulting from blowouts at a given initial density $\rho$ (Figures~\ref{fig:ni} and \ref{fig:ti}), with the caveat that our setup does not account for the density gradient present in the blowout scenario. For a candidate thermonuclear transient resulting from a double-detonation model, a measure of the amount of $^{44}$Ti produced by the event, in combination with a measure of its peak brightness, which can be used to estimate the mass of $^{56}$Ni produced \citep{Ruiter:2013id} \citep[and thus the mass of the C/O core that exploded,][]{Sim:2010hl}, may be able to uniquely constrain the nature of the progenitor system.

\begin{figure}[t]
\centering\includegraphics[width=\linewidth,clip=true]{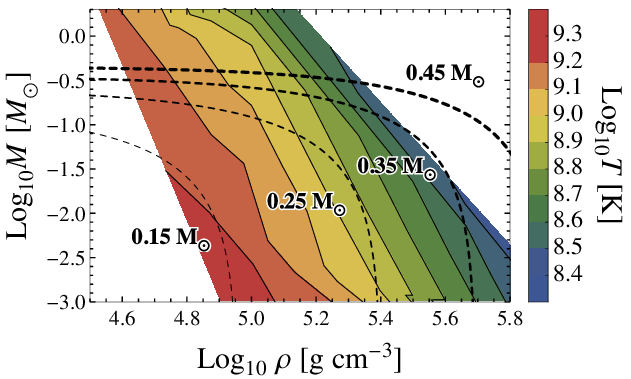}
\caption{Critical masses. The minimal peak temperature required to detonate a given mass $M=\rho\xi_{\rm crit}^3/M_{\odot}$ of density $\rho$ is displayed. Additionally, we have included the trajectories of internal mass as a function of density for $n=3/2$ polytropes with masses $0.15 M_{\odot}$, $0.25 M_{\odot}$, $0.35 M_{\odot}$, and $0.45 M_{\odot}$.}
\label{fig:mass}
\end{figure}

In the density regimes that are expected to realistically facilitate He detonations, intermediate mass elements from $^{40}$Ca to $^{48}$Cr are expected to dominate the nucleosynthetic yields. \citet{Perets:2010fs} reported on SN 2005e, an anomalous low-luminosity supernova in which roughly half of the ejected mass was calcium and significant fraction was radioactive titanium. It was concluded, due to the abundance of helium burning products, that helium burning had occurred in this system. Our results are consistent with this conclusion; the average density of pure He WDs tends to be $\Sim 10^{5}$ g/cm$^3$, which coincides roughly with the peak in calcium production (see Figure~\ref{fig:ti}, top left panel). The numerical investigation of \citet{Waldman:2011ia} into He shell detonations on low-mass WDs could produce large amounts of $^{48}$Cr in fuel densities of $\Sim 5\times 10^{5}$ g cm$^{-3}$, which is in agreement with our findings (Figure~\ref{fig:ti}, bottom left panel). The production of $^{48}$Cr is particularly interesting because the energy release and timescale of its decay chain are comparable to those of $^{56}$Ni. Because of this similarity, equal amounts of $^{48}$Cr and $^{56}$Ni would produce events of near-equal brightness and similar temporal evolution.

In some circumstances, such as collisions and tidal disruptions, the material's temperature is increased in a single dynamical episode to super-virial temperatures, as the source of gravity in these systems is external and can be arbitrarily large. In these cases, the scale of the system need not be comparable to the initiation scale. In Figure~\ref{fig:mass}, we transform the critical sizes presented in Figure~\ref{fig:sizes} into the mass $M$ contained within the critical region, and compare this to the integrated mass corresponding to a density $\rho$ within WDs of different masses. We find that the temperature increase is sufficient to produce detonations in fully-virialized He WDs with masses as low as $0.24 M_{\odot}$, although detonations in lower mass WDs are possible if the material is raised significantly above the virial temperature, as can be the case in either a tidal disruption or in a collision with a more massive (possibly relativistic) compact object. This idea is given some credence by the results of hydrodynamical simulations illustrating that $0.2  M_{\odot}$ He WDs do not detonate after a head-on, equal-mass collision \citep{Rosswog:2009dx}, but {\it do} successfully detonate when colliding with a more massive WD or after being tidally compressed and shock-heated by an IMBH \citep{Rosswog:2009dx,Rosswog:2009gg}. However, because $\xi_{\rm crit}$ is exceeded by the CJ length by a factor of $\Sim 100$, the initiation scales are rarely much smaller than the size of the physical systems themselves, and thus incomplete burning is observed even in systems which are driven to super-virial temperatures \citep[e.g.][]{Rosswog:2008gc}.

While this work presents the first systematic study concerning the length of the temperature gradient necessary to initiate a detonation in a pure He environment, the initial conditions we use are a simplification of the physical problem. Therefore, caution should be exercised in the application of these critical radii, which may deviate somewhat from what we have presented when the simplifying assumptions we have made are relaxed. We do find that the critical radii do not depend sensitively on the functional form of the temperature profile, as illustrated in Figure~\ref{fig:gaussian}, but in reality the density, temperature, and/or composition are unlikely to be homogenous and smooth over the initiation region, as they are assumed to be here. Though we have restricted our study to one-dimensional simulations and optimistically employed a 13-species $\alpha$-chain network, we have demonstrated that the variety of possible outcomes of He detonation is larger than anticipated, and that these outcomes have a diversity of predictive applications to upcoming transient searches. 
 
\acknowledgments 
We thank M. Dan, D. Kasen, S. Rosswog, and D. Townsley for stimulating discussions and fruitful collaboration. We thank F. Timmes for making the equation of state and reaction network used in this work publicly available. We thank the referee for providing thoughtful and constructive comments. We acknowledge support from the David and Lucile Packard Foundation, NSF grant: AST-0847563, the NASA Earth and Space Science Fellowship (J.G.), and the DGAPA-PAPIIT-UNAM grant IA101413-2 (F.D.C.).

\bibliographystyle{apj}
\bibliography{apj-jour,ms}

\end{document}